\documentclass[prd,aps,twocolumn,nofootinbib]{revtex4}

\usepackage{graphicx}

\begin{document}

\title{Revisiting the gravitino dark matter and baryon asymmetry from $Q$-ball decay \\
in gauge mediation}

\author{Shinta Kasuya$^{a,b}$, Masahiro Kawasaki$^{c,d}$ and Masaki Yamada$^c$}

\affiliation{
$^a$ Department of Mathematics and Physics,
     Kanagawa University, Kanagawa 259-1293, Japan\\
$^b$ Max-Planck-Institut f\"ur Kernphysik, PO Box 103980, 69029 Heidelberg, Germany \\
$^c$ Institute for Cosmic Ray Research,
     the University of Tokyo, Chiba 277-8582, Japan\\
$^d$ Kavli Institute for the Physics and Mathematics of the Universe (WPI), 
     Todai Institutes for Advanced Study, the University of Tokyo, Chiba 277-8582, Japan}


\begin{abstract}
We reconsider the $Q$-ball decay and reinvestigate the scenario that the amount of the 
baryons and the gravitino dark matter is naturally explained by the decay of the $Q$ balls in the 
gauge-mediated SUSY breaking. We refine the decay rates into baryons, NLSPs, and gravitinos, and
estimate their branching ratios based on the consideration of Pauli blocking. We obtain
a smaller branching into gravitinos than the previous estimate, and the NLSPs are more 
produced by the $Q$-ball decay. However, the efficient annihilations of NLSPs occur afterward 
so that their abundance does not spoil the successful BBN and they only produce negligible 
amount of the gravitinos to the dark matter density by their decay. In this way, we find that 
the scenario with the direct production of the gravitino dark matter from the $Q$-ball decay 
works naturally. 
\end{abstract}



\maketitle

\section{Introduction}
The abundances of the baryons and dark matter are almost comparable in the present 
universe. It may be understood by the fact that both of them have the same origin: $Q$ balls.
The $Q$ balls are nontopological solitons whose stability is guaranteed by the finite 
charge $Q$ \cite{Coleman}. In supersymmetry (SUSY), the $Q$-ball solitons can actually exist \cite{SUSYQball}, 
and they are produced during the course of
the Affleck-Dine baryogenesis \cite{KuSh,EnMc,KK1,KK2,KK3}, so the charge $Q$
is essentially the baryon number. 

In the gauge-mediated SUSY breaking, very large $Q$ balls are stable against the decay 
into nucleons, the lightest particle with a unit baryon number.\footnote{
Some astrophysical aspect of stable $Q$ balls are found in \cite{stableQball}.} 
On the other hand, if the charge
$Q$ is small enough, the $Q$ balls could decay into nucleons (baryons). In the latter case, 
the gravitinos are also produced by the $Q$-ball decay. The gravitino is the lightest SUSY 
particle in the gauge mediation. Therefore, there may be the way to provide both the baryons
and the dark matter in the universe by the $Q$-ball decay \cite{ShKu,DoMc,KK4,DoMc2}.

In Ref.~\cite{KK4}, two of the present authors investigated the scenario to explain both 
the baryon asymmetry and the dark matter in the universe by the $Q$-ball decay, 
considering the decays into nucleons (quarks), gravitinos, and the next to lightest SUSY 
particles (NLSPs), taking into account the big bang nucleosynthesis (BBN) limits of NLSP 
abundances. Recently in Ref.~\cite{KY}, however, another two of us studied the $Q$-ball 
decay more correctly, and found that the decay rate into gravitinos is much smaller than 
that estimated in Ref.~\cite{KK4}. The essence is to consider the finite size of the $Q$ ball 
which becomes smaller than the penetration length of the decay particle in the actual 
gauge-mediation $Q$ ball. 

In addition, the presence of the main decay channel, which produces quarks, affects the decay
into gravitinos because of the filled-up Fermi sea of quarks. Taking into account of this effect,
we estimate the branching ratio of the gravitino production. 

As we will see, it is natural to achieve the right amounts of the baryon asymmetry and
the gravitino dark matter at the same time through the direct production from the $Q$-ball decay
although we find the branching into gravitinos becomes smaller than the previous estimate in
Ref.\cite{KK4}. In this case, more NLSPs are produced by the $Q$-ball decay, but they 
annihilate afterwards \cite{DoMc2} and the remaining NLSPs provide only the negligible 
amount of the gravitinos to the dark matter density. Moreover, even for $m_{3/2} \gtrsim $ GeV, 
severe BBN constraints  could be avoided if the $Q$-ball decay temperature is high enough.

The structure of the Letter is as follows. In the next section, we review the basic features of
the $Q$ ball in the gauge-mediated SUSY breaking. In Sec.III, we provide the decay rates
of the $Q$ ball, extracted from Ref.~\cite{KY}, and estimate the branching of the decays into NLSPs 
and gravitinos. In Sec.IV, we reestimate the abundances of
the baryons, the gravitinos, and the NLSPs, and seek for the working parameters in Sec.V. 
Here we find that the NLSPs produced by the $Q$-ball decay annihilate efficiently afterward so
that they do not harm the BBN and only produce negligible amount of gravitinos.
We conclude in Sec.VI. In the Appendix A, we show, for completeness, some results concerned 
with the so-called indirect case where the gravitino dark matter is produced by the decay of the 
NLSPs which are created by the $Q$-ball decay.

\section{$Q$ ball in the gauge mediation}
The $Q$ ball is constructed from the scalar field called the flat direction in SUSY
theories. It consists of squarks and sleptons, and the potential is given by \cite{log2}
\begin{equation}
V(\Phi)=\left\{\begin{array}{ll}
m_\phi^2 |\Phi|^2 & (|\Phi| \ll M_S), \\[2mm]
M_F^4 \left(\log\frac{|\Phi|^2}{M_S^2}\right)^2 & (|\Phi| \gg M_S),
\end{array}\right.
\label{pot}
\end{equation}
in the gauge-mediated SUSY breaking\footnote{
$M_F$ and $m_\phi$ are related to the parameters used in Ref.~\cite{KY} as 
$M_F = 2^{-1/4} (m_s M_m/g)^{1/2}$ and $m_\phi=\sqrt{2}m_s$, respectively.
}. Here $m_\phi \sim O$(TeV) is a soft breaking mass, and $M_F$ and $M_S$ are
related to the $F$ and $A$ components of a gauge-singlet chiral multiplet $S$ in the 
messenger, respectively, as
\begin{equation}
M_F^4 \equiv \frac{g^2}{(4\pi)^4}\langle F_A\rangle^2, \qquad M_S\equiv \langle S\rangle,
\end{equation}
where $g$ stands for the gauge coupling in the standard model in general.
$M_F$ ranges as
\begin{equation}
10^3 \ {\rm GeV} \lesssim M_F \lesssim \frac{g^{1/2}}{4\pi}\sqrt{m_{3/2} M_{\rm P}},
\label{MFbound}
\end{equation}
where $m_{3/2}$ is the gravitino mass, and $M_{\rm P} (\approx 2.4 \times 10^{18}$ GeV)
is the reduced Planck scale.

$Q$ balls form from the condensate of the flat direction during its helical motion after
it started rotation when $H\sim M_F^2/\phi_{\rm osc}$, where $\phi_{\rm osc}$ is the amplitude
at the onset of rotation. The charge of the created $Q$ ball is estimated as \cite{KK3}
\begin{equation}
\label{Qform}
Q=\beta\left(\frac{\phi_{\rm osc}}{M_F}\right)^4,
\end{equation}
where $\beta \simeq 6\times 10^{-4}$ for a circular orbit ($\varepsilon=1$), while
$\beta \simeq 6\times 10^{-5}$ for an oblate case ($\varepsilon\lesssim 0.1$). Here
$\varepsilon$ denotes the ellipticity of the field orbit. The charge $Q$ is in fact the $\Phi$-number,
and relates to the baryon number of the $Q$ ball as
\begin{equation}
B=bQ.
\end{equation}
Here we assign the baryon number $b$ to $\Phi$-particle. For example, $b=1/3$ for the 
$udd$ direction. The mass, the size, the rotation speed of the field, and the field amplitude
at the center of the $Q$ ball are expressed in terms of the charge $Q$ respectively as
\begin{eqnarray}
&& M_Q \simeq \frac{4\sqrt{2}\pi}{3} \zeta M_F Q^{3/4}, \\
&& R_Q \simeq \frac{1}{\sqrt{2}} \zeta^{-1} M_F^{-1} Q^{1/4}, \\
&& \omega_Q \simeq \sqrt{2} \pi \zeta M_F Q^{-1/4}, \\
&& \phi_Q \simeq \frac{1}{\sqrt{2}} \zeta M_F Q^{1/4},
\end{eqnarray}
where $\zeta$ is $O(1)$ parameter determined by the fit to numerical calculation 
\cite{HNO, KY}\footnote{The actual value is estimated as $\zeta=2^{1/4}\sqrt{c/\pi}$, where
$c=4.8\log (m_s/\omega_0) + 7.4$ in terms of the parameters used in Ref.~\cite{KY}.
In our successful scenario, typically $c \approx 14$, so that $\zeta \approx 2.5$.}.

The $Q$ ball can decay into those particles kinematically allowed. This condition is
written as $\omega_Q > m_{\rm decay}$. Since we are interested in the production
of baryons, it is rephrased as $\omega_Q > b m_N$, where $m_N \simeq$ 1 GeV is 
the nucleon mass, or $\omega_Q > m_q$ before the confinement, where $m_q$ is the 
quark mass. In this Letter, we adopt the bottom mass so that $m_q = m_b \approx 4.2$ GeV
\footnote{
Results do not change even if we take the top quark mass, since this condition is less restrictive.
See Sec. V.},
although it depends on when the $Q$ balls decay and which flat direction it is, for example. 
This leads to $Q < Q_{\rm D}$ with
\begin{equation}
Q_{\rm D} = 4\pi^4 \zeta^4 \left(\frac{M_F}{m_b}\right)^4.
\label{QD}
\end{equation}
Similarly, the $Q$ ball decays into NLSPs, once the condition 
$\omega_Q > m_{\rm NLSP}$ is met. 
This leads to the charge smaller than $Q_{\rm cr}$, where
\begin{equation}
\label{Qcr}
Q_{\rm cr}  =  4\pi^4 \zeta^4 \left(\frac{M_F}{m_{\rm NLSP}}\right)^4.
\end{equation}

\section{$Q$-ball decay}
The $Q$ ball with baryonic charge with $Q < Q_{\rm D}$ can decay into quarks through the interaction
${\cal L}_{\rm int} = f \phi^* q \eta +$h.c., where $q$ is a quark. Since the quark is a fermion, once the Fermi sea 
is filled, the further decay proceeds only when the fermion escapes from the surface of the $Q$ ball. Thus the 
upper bound of the decay rate is determined by the maximum outgoing flow of the quark \cite{evap}. 
This saturation occurs typically for $f \phi_Q \gtrsim \omega_Q$. For $\omega_Q > M_{\eta}$, the elementary 
decay process  $\phi \rightarrow q + \eta$ is allowed kinematically, and the $Q$-ball decay rate is estimated 
as \cite{KY}\footnote{
The decay rate was first derived in Ref.\cite{evap}. It is different from the present result by factor 2,
since they show the pair production rate in the theory of ${\cal L}_{\rm int} = f \phi^* \chi \chi +$h.c.}.
\begin{equation}
\label{Qdecay_satd}
\Gamma_Q^{\rm (sat,d)} 
\simeq \frac{1}{Q} \frac{\omega_Q^3}{96\pi^2} 4 \pi R_Q^2,
\end{equation}
in the limit of $R_Q\omega_Q \rightarrow \infty$. Here we put the index d, since the elementary 
process is the decay. This occurs for rather smaller charge of the $Q$ ball.

On the other hand, for larger charge when $\omega_Q < M_{\tilde{g}}$, the elementary process of 
the $Q$-ball decay into quarks takes place through heavy gluino (or higgsino) exchange as 
$\phi + \phi \rightarrow q + q$. In this case, the effective coupling is estimated as 
$f_q \simeq M_{\tilde{g}}/\phi_Q$ \cite{KY}, and $f_q \phi_Q  >\omega_Q$ is satisfied, 
so the decay rate is saturated. Since quarks can possess twice larger momenta than 
in the decay case, $\omega_Q$ in the decay rate should be replaced by 
$2\omega_Q$, so that, for the scattering, we have \cite{KY}
\begin{equation}
\label{sat}
\Gamma_Q^{\rm (sat,s)} \simeq \frac{1}{Q} \frac{\omega_Q^3}{12\pi^2} 4\pi R_Q^2.
\end{equation}
Numerical calculation in Ref.~\cite{KY} provides the prefactor, and the decay rate into 
quarks is given by
\begin{equation}
\label{decayquark}
\Gamma_Q^{\rm (q)} \simeq 1.1\, N_q \Gamma_Q^{\rm (sat,s)},
\end{equation}
where $N_q$  is the possible degrees of freedom of quarks, and we set it as $3\times 3\times 2=18$
in the following. This is indeed the main channel of the $Q$-ball decay.

At the same time, the $Q$ ball decays into NLSPs (for $\omega_Q > m_{\rm NLSP}$) and gravitinos. 
The elementary process of the NLSP production is 
$\phi \rightarrow q + \chi$. Since $f_{\rm NLSP} \phi_Q \gg \omega_Q$, where
$f_{\rm NLSP} \simeq g$, Eq.(\ref{Qdecay_satd}) can be applied, which results in
\begin{equation}
\label{decayNLSP}
\Gamma_Q^{\rm (NLSP)} \simeq 1.4\,  \Gamma_Q^{\rm (sat,d)},
\end{equation}
where we include the numerical prefactor derived in Ref.~\cite{KY}. Then the branching ratio
of the NLSP is given by
\begin{equation}
\label{BNLSP}
B_{\rm NLSP} = \frac{\Gamma_Q^{\rm (NLSP)}}{\Gamma_Q^{(q)}}.
\end{equation}

Notice that the decay
$\phi \rightarrow q + \chi$ is not Pauli-blocked by the quarks produced in the main
channel of $\phi + \phi \rightarrow q + q$. This is because the rate of the elementary process
of the decay  $\phi \rightarrow q + \chi$ is much faster than that of the scattering $\phi + \phi \rightarrow q + q$,
so the ``seat" in the phase space $0<p<\omega_Q$, which becomes empty due to the outgoing flow of the 
quark, is quickly occupied by the quark produced in the decay $\phi \rightarrow q + \chi$. On the other hand, 
since there is 8 times larger phase space for the quarks produced by the scattering 
$\phi + \phi \rightarrow q + q$, this main channel is not affected by the decay $\phi \rightarrow q + \chi$, either.

Let us now consider the decay into gravitinos. The elementary process is 
$\phi \rightarrow q + \psi_{3/2}$. Since the effective coupling is estimated as
\begin{equation}
f_{3/2} \simeq \frac{\omega_Q^2}{\sqrt{3} m_{3/2} M_{\rm P}},
\end{equation}
it is typically $f_{3/2} \phi_Q \ll \omega_Q$, and the Fermi sea will not be filled by this process alone.
In this case, the $Q$-ball decay rate is estimated as\cite{KY}
\begin{equation}
\label{grav}
\Gamma_Q^{(3/2)} \simeq 
0.9 \left(\frac{\omega_Q\phi_Q}{\langle F \rangle}\right)^2
\Gamma_Q^{\rm (sat,d)},
\end{equation}
where $\langle F \rangle=\sqrt{3}m_{3/2} M_{\rm P}$ is the SUSY breaking $F$ term.
The physical meaning can be understood as follows \cite{KY}. The interaction region is given by 
the penetration length of produced particles $\sim (f \phi_Q)^{-1}$. 
Since the size of the $Q$ ball is $R_Q \sim \omega_Q^{-1}$, so 
that the penetration length of the produced particles becomes larger than the $Q$-ball size: 
$(f \phi_Q)^{-1} \gg \omega_Q^{-1} \sim R_Q$. Then the effective volume of the interaction region
should be the whole volume of the $Q$ ball. 
Therefore, the charge decreasing rate ($dQ/dt$) can be estimated as the product of one-particle decay rate 
($\sim f^2 \omega_Q$),  particle density ($\sim \omega_Q \phi_Q^2$), and the effective
volume ($\sim R_Q^3$).\footnote{
It was first estimated in Ref.\cite{evap} as
$\Gamma_Q \simeq 3\pi \frac{f \phi_Q}{\omega_Q} \Gamma_Q^{\rm (sat)}$ for $f \phi_Q \ll \omega_Q$ and
$R_Q\omega_Q \rightarrow \infty$. This is valid when the penetration length of produced particles 
$\sim (f \phi_Q)^{-1}$ is smaller than the size of the $Q$ ball \cite{KY}.}

However, for $\omega_Q < M_{\tilde{g}}$ where $\phi + \phi \rightarrow q + q$ process is the main channel
of the $Q$-ball decay, the phase spaces of all the quarks are filled by this process, and the rate of the 
elementary process $\phi \rightarrow q + \psi_{3/2}$ is much lower than that of the main process. In this case,
the decay into gravitinos is determined by the probability to occupy the ``seat" in phase space of the each 
quark by the process $\phi \rightarrow q + \psi_{3/2}$. Then the branching ratio of the gravitino production 
by the $Q$-ball decay should be the ratio of the rate for the elementary processes. 
It can be understood as follows. The quark produced by a single reaction of $\phi \rightarrow q + \psi_{3/2}$ 
have momentum with $0<p<\omega_Q$ so that it occupies one ``seat" in the phase space of $0<p<\omega_Q$.
On the other hand, the process $\phi + \phi \rightarrow q + q$ produces one quark with $0<p<\omega_Q$
and the other quark with $\omega_Q<p<2\omega_Q$. Thus, one ``seat"  in the phase space of 
$0<p<\omega_Q$ is occupied by a single $\phi + \phi \rightarrow q + q$ reaction.
Thus, we have
\begin{equation}
\label{Bgrav}
B_{3/2} \simeq \frac{\Gamma(\phi\rightarrow q \psi_{3/2})}{\Gamma(\phi\phi\rightarrow qq)}
= \frac{f_{3/2}^2}{f_q^2}.
\end{equation}

Notice that the process $\phi \rightarrow q + \chi$ does not affect the process $\phi \rightarrow q + \psi_{3/2}$.
This is because the former process cannot fill all the empty ``seats" in the phase spaces of $N_q$ quarks,
since the Pauli blocking of the NLSP prevents to produce multiple quarks at one time. On the other hand,
there is no Pauli blocking of the gravitino, because $f_{3/2}\phi_Q \ll \omega_Q$. Therefore, the latter process
can fill the empty ``seats" of all the quarks, although it is suppressed by the main channel of quark production as
mentioned above.

For $\omega_Q > M_{\tilde{g}}$, the process $\phi \rightarrow q + \tilde{g}$ is the main decay channel.
In this case, the decay rate of the $Q$ ball into quarks is obtained as
\begin{equation}
\Gamma_Q^{(q)} \simeq 1.4 N_S \Gamma_Q^{\rm (sat, d)},
\end{equation}
where $N_S$ is the number of the gluinos (or more precisely, the number of the sparticles with the mass
smaller than $\omega_Q$ which has coupling to $\phi$). If $N_q > N_S$, not all the ``seats" in the phase 
spaces of the quarks are filled by this process. Thus, the process $\phi \rightarrow q + \psi_{3/2}$ is not
affected by this main channel. Therefore, Eq.(\ref{grav}) is applicable for the rate of $Q$-ball decay into gravitinos,
and the branching ratio is estimated as
\begin{equation}
B_{3/2} \simeq \frac{\Gamma_Q^{(3/2)}}{N_S \Gamma_Q^{\rm (sat,d)}}.
\end{equation}

\section{Abundances}
Now we have in our hand the revised decay rate into baryons (\ref{decayquark}) and the branching into
gravitinos (\ref{Bgrav}). Since the main decay channel is the decay into baryons, the temperature 
at the decay is given by
\begin{eqnarray}
T_{\rm D} & = & \left(\frac{90}{4\pi^2 N_{\rm D}}\right)^{1/4}\sqrt{\Gamma_Q^{\rm (q)} M_{\rm P}}
\nonumber \\
& \simeq & 0.58 \ {\rm GeV} \left(\frac{\zeta}{2.5}\right)^{1/2} 
\left(\frac{N_q}{18}\right)^{1/2} \left(\frac{N_{\rm D}}{61.75}\right)^{-1/4}
\nonumber \\ & & \hspace{10mm} \times
\left(\frac{M_F}{10^7 \ {\rm GeV}}\right)^{1/2} 
\left(\frac{Q}{10^{22}}\right)^{-5/8},
\end{eqnarray}
where $N_{\rm D}$ is the relativistic degrees of freedom at the decay time.

We have the relations between number densities of baryons, gravitinos, and NLSPs, and 
that of the $\phi$ field as \cite{KK4}
\begin{eqnarray}
\label{nb}
& & n_b \simeq \varepsilon b n_\phi, \\[3mm]
& & n_{3/2} \simeq B_{3/2} n_\phi, \\[2mm]
& & n_{\rm NLSP} \simeq B_{\rm NLSP} \frac{Q_{\rm cr}} {Q} n_\phi,
\label{nNLSP}
\end{eqnarray}
where $B_{3/2}$ is the branching ratio of the decay into gravitinos Eq.(\ref{Bgrav}), and
$B_{\rm NLSP}$ is the branching into NLSPs Eq.(\ref{BNLSP}), once it is allowed kinematically 
when $Q<Q_{\rm cr}$. If the charge $Q$ is smaller than $Q_{\rm cr}$
when the $Q$ ball formed, one should set $Q_{\rm cr}/Q=1$ in Eq.(\ref{nNLSP}).

The actual abundances are dependent on whether or not $Q$ balls 
dominate the energy density of the universe before the decay. The baryon abundances 
for the $Q$-ball dominated (QD) and non-$Q$-ball dominated (NQD) 
cases are thus given by
\begin{equation}
\label{Yb}
Y_b \equiv \frac{n_b}{s} = \left\{\begin{array}{l}
\displaystyle{
\frac{3T_{\rm D}}{4} \left.\frac{n_b}{\rho_Q}\right|_{\rm D}
= \frac{3T_{\rm D}}{4} \left.\frac{n_b}{\rho_Q}\right|_{\rm osc}}\\[4mm]
\displaystyle{\hspace{8mm}
\simeq \frac{3T_{\rm D}}{4} \frac{\varepsilon b n_\phi}{\frac{4}{3}\omega_Q n_\phi}
\simeq \frac{9T_{\rm D}\varepsilon b}{16\omega_Q}, } \qquad {\rm (QD)} \\[5mm]
\displaystyle{
\frac{3T_{\rm RH}}{4} \left.\frac{n_b}{\rho_{\rm rad}}\right|_{\rm RH} 
= \frac{3T_{\rm RH}}{4} \left.\frac{n_b}{\rho_{\rm inf}}\right|_{\rm osc}}\\[4mm]
\hspace{8mm}\displaystyle{
\simeq  \frac{9}{8\sqrt{2}} \varepsilon b \beta^{-3/4}\frac{M_F T_{\rm RH}}{M_{\rm P}^2} Q^{3/4},}
\ {\rm (NQD)} 
\end{array}\right.
\label{Yb_org}
\end{equation}
where Eq.(\ref{nb}) and
\begin{equation}
\label{rhoQ}
\rho_Q \simeq M_Q n_Q \simeq M_Q \frac{n_\phi}{Q} \simeq \frac{4}{3} \omega_Q n_\phi,
\end{equation}
are used in the QD case, while $n_\phi \simeq m_{\rm eff} \phi_{\rm osc}^2$, 
$m_{\rm eff} = \sqrt{V''}\simeq 2\sqrt{2}M_F^2/\phi_{\rm osc}$, 
$3H_{\rm osc} = m_{\rm eff}$, and Eq.(\ref{Qform}) are used in the NQD case. 
On the other hand, the abundance of the gravitino dark matter is determined by the ratio with
the baryon density as
\begin{equation}
\label{BDM}
5\approx \frac{\rho_{3/2}}{\rho_b} = \frac{m_{3/2}}{m_N}\frac{n_{3/2}}{n_b} \simeq
\displaystyle{\frac{m_{3/2}}{m_N} \frac{B_{3/2}}{\varepsilon b}}.
\end{equation}
Therefore, we can eliminate $\varepsilon b$ in Eq.(\ref{Yb}) using Eq.(\ref{BDM}) to obtain 
the baryon abundance as, setting $\beta=6\times 10^{-5}$, 
\begin{eqnarray}
& & \hspace{0mm}
\frac{Y_b^{\rm (QD)}}{10^{-10}} \simeq 5.3 \times 10^{-5}
\left(\frac{\zeta}{2.5}\right)^{11/2} 
\left(\frac{N_q}{18}\right)^{1/2}
\nonumber \\ & & \hspace{2mm} \times 
\left(\frac{N_{\rm D}}{61.75}\right)^{-1/4}
\left(\frac{m_{3/2}}{100 \ {\rm MeV}}\right)^{-1}
\left(\frac{M_{\tilde{g}}}{3 \ {\rm TeV}}\right)^{-2}
\nonumber \\ & & \hspace{7mm} \times 
\left(\frac{M_F}{10^7 \ {\rm GeV}}\right)^{11/2}
\left(\frac{Q}{10^{22}}\right)^{-7/8},
\label{Yb_QDdir}
\end{eqnarray}
and 
\begin{eqnarray}
\frac{Y_b^{\rm (NQD)}}{10^{-10}} & \simeq & 3.9 \times 10^{-5}
\left(\frac{\zeta}{2.5}\right)^6 \left(\frac{T_{\rm RH}}{10^4 \ {\rm GeV}}\right)
\nonumber \\ & & \hspace{0mm} \times
\left(\frac{m_{3/2}}{100 \ {\rm MeV}}\right)^{-1}
\left(\frac{M_{\tilde{g}}}{3 \ {\rm TeV}}\right)^{-2}
\nonumber \\ & & \hspace{0mm} \times
\left(\frac{M_F}{10^7 \ {\rm GeV}}\right)^7
\left(\frac{Q}{10^{22}}\right)^{1/4},
\label{Yb_NQDdir}
\end{eqnarray}
for the $Q$-ball dominance and non-dominance, respectively. These relations must be hold
for simultaneous explanation of the baryon and gravitino dark matter abundances.

\begin{figure}[ht!]
\includegraphics[width=90mm]{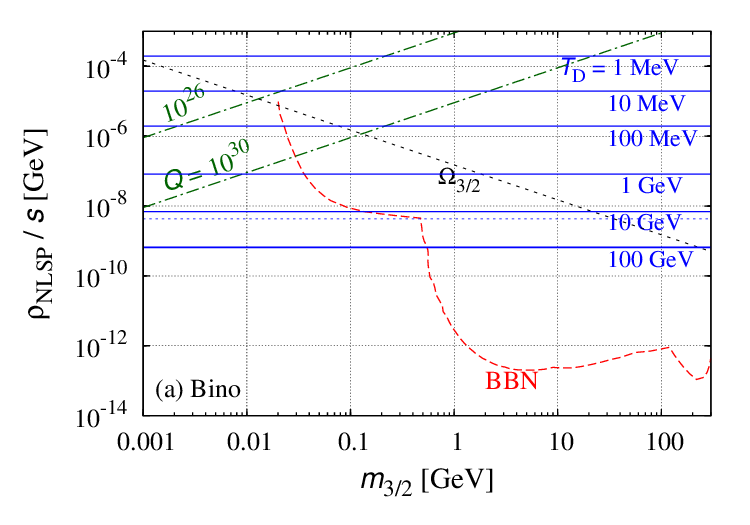}
\includegraphics[width=90mm]{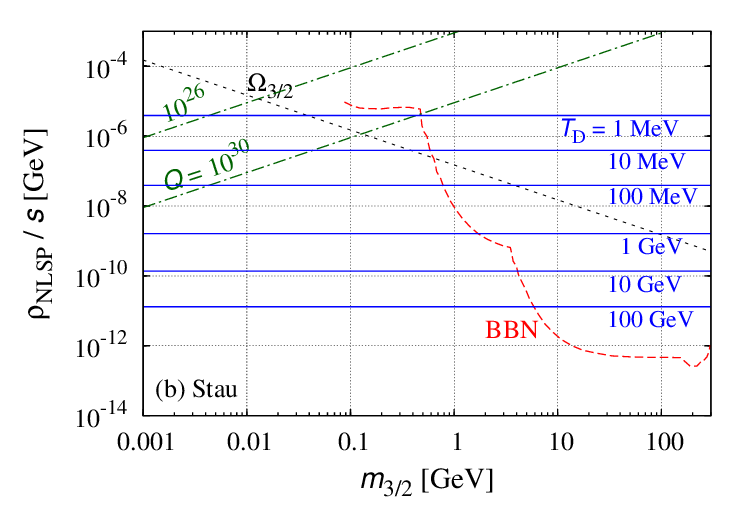}
\includegraphics[width=90mm]{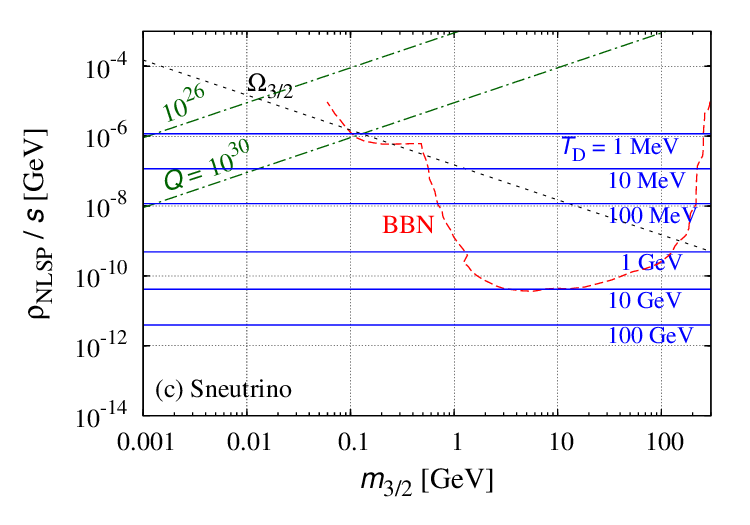}
\caption{NLSP abundances produced by the $Q$-ball decay for $Q=10^{26}$ 
and $10^{30}$ in dark green dotted-dashed lines. The abundances fixed by the annihilation
are denoted by blue solid lines. Black dotted line is the abundance that the NLSP decay gives
the right amount of the gravitino dark matter. Above red dashed lines are excluded by the BBN 
constraints \cite{BBN}.}
\label{fig_NLSP}
\end{figure}

The NLSP abundance produced by the $Q$-ball decay is estimated as
\begin{eqnarray}
\frac{\rho_{\rm NLSP}^{(Q)}}{s} & = & m_{3/2} Y_{3/2} \frac{\rho_{\rm NLSP}}{\rho_{3/2}}
\simeq 5 m_N Y_b \frac{m_{\rm NLSP}}{m_{3/2}} \frac{n_{\rm NLSP}}{n_{3/2}}
\nonumber \\ & \simeq &
5 m_N Y_b \frac{m_{\rm NLSP}}{m_{3/2}} \frac{B_{\rm NLSP}Q_{\rm cr}}{B_{3/2}Q}
\nonumber \\ & \simeq &
8.1\times 10^{-2} \ {\rm GeV} \left(\frac{Y_b}{10^{-10}}\right) \left(\frac{N_q}{18}\right)^{-1}
\left(\frac{\zeta}{2.5}\right)^{-2} 
\nonumber \\ & & \hspace{1mm} \times 
\left(\frac{m_{3/2}}{100 \ {\rm MeV}}\right) \left(\frac{M_{\tilde{g}}}{3 \ {\rm TeV}}\right)^2
\left(\frac{m_{\rm NLSP}}{300 \ {\rm GeV}}\right)^{-3}
\nonumber \\ & & \hspace{1mm} \times 
\left(\frac{M_F}{10^7 \ {\rm GeV}}\right)^{-2}
\left(\frac{Q}{10^{22}}\right)^{-1/2},
\label{NLSPabundance}
\end{eqnarray}
at the decay time. Here and hereafter we take $m_{\rm NLSP}=300$ GeV, although we keep
showing its dependence.
It is shown for some fixed value of the charge $Q$ with $M_F=10^7$ GeV in the dark green dotted-dashed
lines in Fig.~\ref{fig_NLSP}. 

The NLSP abundance is usually determined by the subsequent annihilations of NLSPs 
as $n_{\rm NLSP}^{\rm (ann)} \simeq H(T_{\rm D})/\langle\sigma v\rangle$ \cite{DoMc2}.
The annihilation cross sections for the bino, stau, and sneutrino NLSP are estimated 
as \cite{FeSuTa}
\begin{eqnarray}
\langle\sigma v\rangle_{\tilde{B}} & \simeq & 
1.5\times 10^{-10} {\rm GeV}^{-2} \left(\frac{m_{\rm NLSP}}{300 \ {\rm GeV}}\right)^{-2}, \\
\langle\sigma v\rangle_{\tilde{\tau}} & \simeq & 
7.4\times 10^{-9} {\rm GeV}^{-2} \left(\frac{m_{\rm NLSP}}{300 \ {\rm GeV}}\right)^{-2}, \\
\langle\sigma v\rangle_{\tilde{\nu}} & \simeq & 
2.5\times 10^{-8} {\rm GeV}^{-2} \left(\frac{m_{\rm NLSP}}{300 \ {\rm GeV}}\right)^{-2}.
\end{eqnarray}
They lead to
\begin{eqnarray}
\label{rho_bino_ann}
\left.\frac{\rho_{\rm NLSP}^{\rm (ann)}}{s}\right|_{\tilde{B}} 
& \simeq & 2.0 \times 10^{-4} \ {\rm GeV}
\left(\frac{N_{\rm D}}{10.75}\right)^{-1/2} 
\nonumber \\ & & \hspace{5mm} \times
\left(\frac{m_{\rm NLSP}}{300 \ {\rm GeV}}\right)^3 
\left(\frac{T_{\rm D}}{\rm MeV}\right)^{-1}, \\
\label{rho_stau_ann}
\left.\frac{\rho_{\rm NLSP}^{\rm (ann)}}{s}\right|_{\tilde{\tau}} 
& \simeq & 3.9 \times 10^{-6} \ {\rm GeV}
\left(\frac{N_{\rm D}}{10.75}\right)^{-1/2} 
\nonumber \\ & & \hspace{5mm} \times
\left(\frac{m_{\rm NLSP}}{300 \ {\rm GeV}}\right)^3 
\left(\frac{T_{\rm D}}{\rm MeV}\right)^{-1}, \\
\label{rho_snu_ann}
\left.\frac{\rho_{\rm NLSP}^{\rm (ann)}}{s}\right|_{\tilde{\nu}} 
& \simeq & 1.2 \times 10^{-6} \ {\rm GeV}
\left(\frac{N_{\rm D}}{10.75}\right)^{-1/2} 
\nonumber \\ & & \hspace{5mm} \times
\left(\frac{m_{\rm NLSP}}{300 \ {\rm GeV}}\right)^3 
\left(\frac{T_{\rm D}}{\rm MeV}\right)^{-1}. 
\end{eqnarray}
When the density of the NLSPs produced from $Q$-ball decay is larger than 
the annihilation density, $\rho_{\rm NLSP}^{(Q)} > \rho_{\rm NLSP}^{\rm (ann)}$, 
the NLSP density will settle down to the annihilation value in the end. The annihilation
abundances are displayed for several different decay temperatures $T_{\rm D}$ in blue 
solid lines in Fig.~\ref{fig_NLSP}.

The abundance is generally limited by the BBN bound for each species of NLSPs \cite{BBN}, 
shown in red dashed lines, or bounded by the NLSP abundance that accounts the right amount 
of the gravitino dark matter by the NLSP decay, denoted by black dotted line. 
The latter limit is given by \cite{BBN, KK4}
\begin{eqnarray}
\left.\frac{\rho_{\rm NLSP}}{s}\right|_{\Omega_{3/2}} & = & 1.5 \times 10^{-6} \ {\rm GeV}
\left(\frac{m_{\rm NLSP}}{300 \ {\rm GeV}}\right) 
\nonumber \\ & &
\times \left(\frac{m_{3/2}}{100 \ {\rm MeV}}\right)^{-1}
\left(\frac{Y_b}{10^{-10}}\right).
\label{NLSP_Omega}
\end{eqnarray}
As we see later, the $Q$-ball charge is $\lesssim 10^{26}$, so that the NLSP abundance
is fixed by the annihilation density. Therefore, for larger decay temperature $T_{\rm D}$, which
will be achieved in the successful scenario, the gravitinos produced by the decay of
NLSPs can entirely be negligible for the gravitino dark matter abundance. Moreover,
severe BBN constraints are evaded, and we can take larger gravitino mass $m_{3/2}$
for successful scenarios.

\section{Constraining model parameters}
\begin{figure}[h!]
\includegraphics[width=90mm]{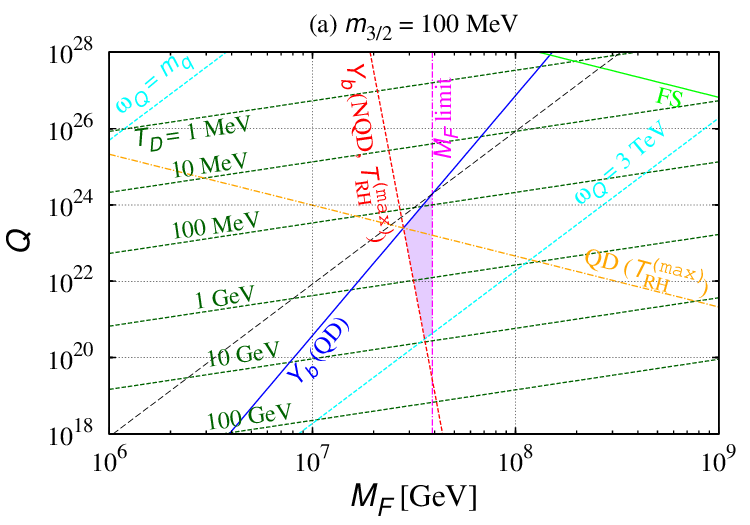}
\includegraphics[width=90mm]{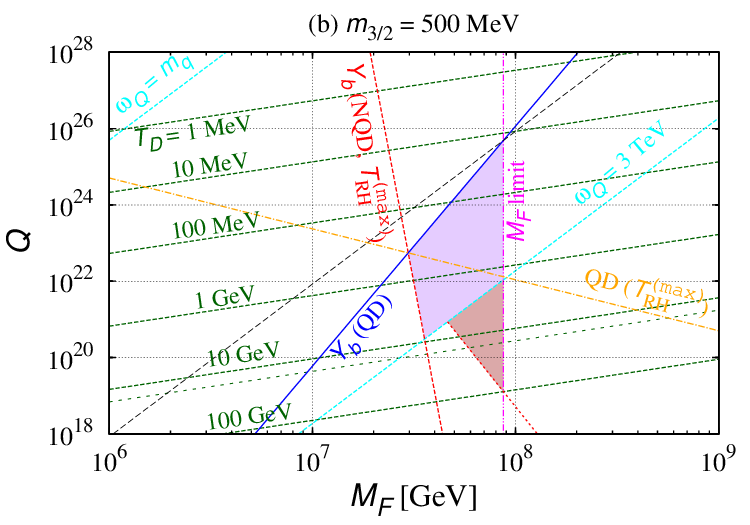}
\caption{Allowed region for $m_{3/2}=$ (a) 100 MeV and (b) 500 MeV, which is surrounded 
by the $Y_b$ line for QD case (blue solid line) and for NQD case with $T_{\rm RH}$ being fixed
(red dashed line), and by $M_F$ limit (magenta double-dotted dashed line).
Iso-$T_D$ contours are shown in dark green lines. We also show the kinematical limit (cyan dashed)
and the free streaming constraint (green solid). Black long dashed line represents the trace of
the upper limit of the allowed region when $m_{3/2}$ changes.}
\label{fig_QMF}
\end{figure}

Our scenario should simultaneously explain the right amount of the baryon asymmetry 
and the gravitino dark matter both from the $Q$-ball decay. Since $Y_b \simeq 10^{-10}$, 
for successful scenario, we can obtain, from Eqs.(\ref{Yb_QDdir}) and (\ref{Yb_NQDdir}),
the relations that hold for the charge $Q$ as
\begin{eqnarray}
\label{QYb_QDdir}
Q & \simeq & 3.8 \times 10^{20} \left(\frac{\zeta}{2.5}\right)^{44/7} \left(\frac{N_q}{18}\right)^{4/7} 
\nonumber \\ & & \hspace{8mm} \times
 \left(\frac{N_{\rm D}}{61.75}\right)^{-2/7}
\left(\frac{m_{3/2}}{100 \ {\rm MeV}}\right)^{-8/7}
\nonumber \\ & & \hspace{8mm} \times
\left(\frac{M_{\tilde{g}}}{3 \ {\rm TeV}}\right)^{-16/7}
\left(\frac{M_F}{10^7\ {\rm GeV}}\right)^{44/7}, \\
Q & \simeq & 4.2 \times 10^{39} \left(\frac{\zeta}{2.5}\right)^{-24}
\left(\frac{T_{\rm RH}}{10^4 \ {\rm GeV}}\right)^{-4}
\left(\frac{m_{3/2}}{100 \ {\rm MeV}}\right)^4
\nonumber \\ & & \hspace{18mm} \times
\left(\frac{M_{\tilde{g}}}{3 \ {\rm TeV}}\right)^8
\left(\frac{M_F}{10^7\ {\rm GeV}}\right)^{-28},
\label{QYb_NQDdir}
\end{eqnarray}
for QD and NQD cases, respectively. The latter depends on the reheating 
temperature after inflation, $T_{\rm RH}$. It has an upper bound so that thermally 
produced gravitinos do not overdominate the universe, which results in \cite{KTY}
\begin{equation}
T_{\rm RH} \lesssim 8.3 \times 10^4 \ {\rm GeV} \left(\frac{m_{3/2}}{100 \ {\rm MeV}}\right)
\left(\frac{M_{\tilde{g}}}{3 \ {\rm TeV}}\right)^{-2}.
\label{TRHcond}
\end{equation}
Taking account of this upper bound in Eq.(\ref{QYb_NQDdir}), we obtain the lower limit as
\begin{equation}
Q \simeq 8.8\times 10^{35} \left(\frac{\zeta}{2.5}\right)^{-24}
\left(\frac{M_{\tilde{g}}}{3 \ {\rm TeV}}\right)^{16}
\left(\frac{M_F}{10^7\ {\rm GeV}}\right)^{-28}.
\label{QYb_NQDdir2}
\end{equation}
Notice that it is independent of $m_{3/2}$. Eqs.(\ref{QYb_QDdir}) and (\ref{QYb_NQDdir2})
are shown in blue solid and red dashed lines, respectively, in Fig.~\ref{fig_QMF}.

Together with the upper limit (\ref{MFbound}) on $M_F$, shown in magenta double-dotted 
dashed lines in Fig.~\ref{fig_QMF}, we have the allowed regions in the parameter 
space ($M_F$, $Q$). We show in Fig.~\ref{fig_QMF} two cases with different 
gravitino masses: (a) 100 MeV and (b) 500 MeV.

The boundary of the $Q$-ball domination/non-domination is derived from the relation
\begin{eqnarray}
1 & = &\left.\frac{\rho_Q}{\rho_r}\right|_{\rm D} 
=\left.\frac{\rho_Q}{\rho_r}\right|_{\rm RH}\frac{T_{\rm RH}}{T_{\rm D}}
=\left.\frac{\rho_Q}{\rho_{\rm inf}}\right|_{\rm osc}\frac{T_{\rm RH}}{T_{\rm D}}
\nonumber \\
& = & \left.\frac{n_Q}{n_{\rm inf}}\right|_{\rm osc}\frac{T_{\rm RH}}{T_{\rm D}}
= \frac{Y_b^{\rm (NQD)}}{Y_b^{\rm (QD)}}
\nonumber \\
& \simeq & 6.9 \times 10^{-4} \left(\frac{\zeta}{2.5}\right)^{1/2}
\left(\frac{N_q}{18}\right)^{-1/2}
\left(\frac{N_{\rm D}}{61.75}\right)^{1/4}
\nonumber \\ & & \times
\left(\frac{T_{\rm RH}}{10^4 \ {\rm GeV}}\right) 
\left(\frac{M_F}{10^7 \ {\rm GeV}}\right)^{3/2}
\left(\frac{Q}{10^{22}}\right)^{9/8},
\end{eqnarray}
where we use Eq.(\ref{Yb_org}). This leads to the following boundary of the $Q$-ball 
dominance on ($M_F$,$Q$) parameter space, given by 
\begin{eqnarray}
Q & \simeq & 6.5 \times 10^{24} \left(\frac{\zeta}{2.5}\right)^{-4/9}
\left(\frac{N_q}{18}\right)^{4/9}
\left(\frac{N_{\rm D}}{61.75}\right)^{-2/9}
\nonumber \\ & & \hspace{5mm}\times
\left(\frac{T_{\rm RH}}{10^4\ {\rm GeV}}\right)^{-8/9}
\left(\frac{M_F}{10^7\ {\rm GeV}}\right)^{-4/3},
\label{Qdomi}
\end{eqnarray}
above (below) which corresponds to the (N)QD case, shown in orange 
dotted-dashed lines in Fig.~\ref{fig_QMF}.@As $T_{\rm RH}$ decreases to lower values, the baryon 
abundance line in NQD case sweeps the hatched region. 

We cut off the region at $\omega_Q =M_{\tilde{g}}=$ 3 TeV in Fig.~\ref{fig_QMF}. In the lower region, 
the decay process $\phi \rightarrow q \tilde{g}$ is kinematically allowed. In this case, this process could be
the main decay channel of the $Q$ ball, and we show the lower limit in red dotted line below 
$\omega_Q =$ 3 TeV line in Fig.~\ref{fig_QMF}(b), where the allowed region is shown in brown hatched
region. However, depending on the mass spectrum, $\phi +\phi\rightarrow q+q$ by higgsino exchange could
be the main decay channel of the $Q$ ball, or other decay processes $\phi\rightarrow q+X$ (X: other sparticles) 
could also contribute to the $Q$-ball decay into quarks, which is highly model dependent.

The allowed region for $m_{3/2} = 100$ MeV shown in Fig.~\ref{fig_QMF}(a) applies to
stau and sneutrino NLSPs, while it does not to the bino, as can be seen from Fig.~\ref{fig_NLSP}
that the decay temperature should be $\gtrsim 10$ GeV for the bino NLSP in order to
evade the BBN constraint, while $\gtrsim$ a few MeV is possible for stau and sneutrino NLSP.
Similarly, the allowed region for $m_{3/2} = 500$ MeV is applicable to stau and sneutrino cases,
while the bino case is allowed only for $T_D \gtrsim 16$ GeV (below the dark green dotted line). 
(See also Fig.~\ref{fig_NLSP}(a) where we show $T_D \simeq 16$ GeV in dotted line.)

Black long dashed lines show the trace of the upper bound of allowed regions when
$m_{3/2}$ is changed. Therefore, the whole allowed region 
is below this long dashed line and above red short dashed line, and the lower limit of the
gravitino mass is obtained as $m_{3/2} \gtrsim$ 50 MeV. On the other hand, one can take larger
gravitino mass of $O$(GeV) in the stau case, and even much larger in the sneutrino case,
as can be seen from Fig.~\ref{fig_NLSP}. It is allowed typically for $T_{\rm D} \gtrsim 10$ GeV,
which corresponds to the lower part in ($M_F$, $Q$) plane. It will eventually be constrained
by the fact that such small charge $Q$ balls may evaporate in the thermal bath, or 
the $Q$ ball cannot be the gauge-mediation type, where the logarithmic term can
no longer dominate over the mass term (see Eq.(\ref{pot})).

Now let us check that the allowed regions obtained above are consistent with more general 
conditions, which indeed turn out that they are less restrictive. We show the decay temperature 
$T_D$ in dark green lines, the kinematical bound $\omega_Q > m_q$ 
in cyan lines, and the free streaming limit in green lines in Fig.~\ref{fig_QMF}.

First we see the constraint on the decay temperature $T_{\rm D}$.
Since the decay of the $Q$ ball should take place before the BBN, the decay temperature
must be larger than $\sim$MeV. The $T_{\rm D}$ contour is given by
\begin{eqnarray}
Q & \simeq & 5.3 \times 10^{26} \left(\frac{\zeta}{2.5}\right)^{4/5}
\left(\frac{N_q}{18}\right)^{4/5} \left(\frac{N_{\rm D}}{10.75}\right)^{-2/5}
\nonumber \\ & & \hspace{10mm} \times
\left(\frac{T_{\rm D}}{\rm MeV}\right)^{-8/5}
\left(\frac{M_F}{10^7 \ {\rm GeV}}\right)^{4/5}.
\label{TDconst}
\end{eqnarray}

Next we consider the kinematical bound. The $Q$ ball is kinematically allowed to 
decay into quarks. This is achieved for $Q < Q_{\rm D}$ (see Eq.(\ref{QD})):
\begin{equation}
Q < 5.0 \times 10^{29} \left(\frac{\zeta}{2.5}\right)^4
\left(\frac{M_F}{10^7 \ {\rm GeV}}\right)^4.
\end{equation}

In addition, since the gravitino mass is relatively small, the produced 
gravitinos could free-stream to affect large scale structures. In order to avoid such a case,
we impose that the free streaming length should not exceed $\sim$ Mpc \cite{FSbib}, which
results in the bound on the present-day free streaming velocity as $v_0 \lesssim 3.7 \times 10^{-7}$.
$v_0$ is estimated as
\begin{equation}
v_0 \simeq \frac{1}{2} \frac{\omega_Q}{m_{3/2}}\frac{T_0}{T_{\rm D}}
\left(\frac{N_0}{N_{\rm D}}\right)^{1/3},
\end{equation}
where $T_0$ and $N_0$ are respectively the temperature and the relativistic degrees of freedom
at present. Therefore, the constraint is obtained as
\begin{eqnarray}
Q & \lesssim & 3.1 \times 10^{29} \left(\frac{\zeta}{2.5}\right)^{-4/3}
\left(\frac{N_q}{18}\right)^{4/3} \left(\frac{N_{\rm D}}{10.75}\right)^{2/9}
\nonumber \\ & & \hspace{5mm} \times
\left(\frac{m_{3/2}}{\rm 100 \ MeV}\right)^{8/3}
\left(\frac{M_F}{10^7 \ {\rm GeV}}\right)^{-4/3}.
\label{FSdir}
\end{eqnarray}

\section{Conclusions}
We have reconsidered the decay rates of the $Q$ ball, and reinvestigated the scenario that the 
amount of the baryons and the gravitino dark matter is naturally explained by the decay of the 
$Q$ balls in the gauge-mediated SUSY breaking. Taking into account the more correct 
decay rates into baryons derived recently in Ref.~\cite{KY},  the branching into NLSPs and 
gravitinos based on consideration of the Pauli blocking, and also the $Q$-ball domination 
and/or the free streaming condition, we have found that the scenario actually works.
Although the branching into gravitinos is smaller than the previous estimate, which
leads to the more relative production of NLSPs by the decay of the $Q$ balls, the actual 
abundance of the NLSPs is suppressed by the annihilation effects. 

We have found that the scenario works typically for $m_{3/2} = 50$ MeV $-$ 5 GeV,
$M_F = 3 \times 10^7 -  3\times 10^8$ GeV, and $Q = 10^{19} - 10^{26}$, 
although there are some differences among the NLSP species. Therefore,
it is natural to have this scenario of simultaneous direct creation of the baryon asymmetry
and the gravitino dark matter from the $Q$-ball decay.

\section*{Acknowledgments}
S.K. would like to thank Particle and Astroparticle Division of Max-Planck-Institut f\"ur
Kernphysik, Heidelberg for hospitality.
The work is supported by Grant-in-Aid for Scientific Research  
23740206 (S.K.), 22540267 (M.K.) and 21111006 (M.K.) 
from the Ministry of Education, Culture, Sports, Science and 
Technology in Japan, and also by World Premier International Research 
Center Initiative (WPI Initiative), MEXT, Japan.

\appendix
\section{Indirect production of the gravitino dark matter via NLSPs}
We briefly summarize the case of the indirect production of the gravitino dark matter 
from the $Q$-ball decay. The $Q$ ball first decays into NLSPs, which eventually decay
into the gravitinos that become the dark matter of the universe. In actual scenario,
there are two cases. The one is that the NLSPs annihilate to reduce their abundance to
the annihilation density (Eqs.(\ref{rho_bino_ann}) $-$ (\ref{rho_snu_ann})). This is essentially
the same case considered in Ref.\cite{DoMc2}.
The other is the case that the amount of the NLSPs produced by the $Q$-ball decay is small 
enough that they do not annihilate to reduce the amount. In either case, the NLSP abundance
should coincide with Eq.(\ref{NLSP_Omega}).

In the former case, where the annihilation is effective, it can be readily read off 
from Fig.~\ref{fig_NLSP} for the allowed range of the gravitino mass $m_{3/2}$ (or equivalently,
the decay temperature of the $Q$ ball, $T_{\rm D}$): 
\begin{equation}
\left\{\begin{array}{l}
\begin{array}{l}
m_{3/2}  =  2.8 - 20 \ {\rm MeV} \\
(T_{\rm D}  =  3.7 - 26 \ {\rm MeV}),
\end{array}  \hspace{11mm} {\rm (bino)} \\[4mm]
\begin{array}{l}
m_{3/2}  =  39 - 500 \ {\rm MeV} \\
(T_{\rm D}  =  1 - 13 \ {\rm MeV}),
\end{array}  \hspace{10mm}{\rm (stau)} \\[4mm]
\begin{array}{l}
m_{3/2}  =  (3-5)\times 100 \ {\rm MeV} \\
(T_{\rm D}  =  1.6 - 3.9 \ {\rm MeV}).
\end{array}  \hspace{3mm} {\rm (sneutrino)} 
\end{array}\right.
\label{TD_bound}
\end{equation}
Notice that the lower bound in the bino NLSP is determined by 
$\Gamma_Q^{(q)} > \Gamma_{\rm NLSP}$, where the NLSP decay rate is estimated as
\begin{equation}
\Gamma_{\rm NLSP} = \frac{m_{\rm NLSP}^5}{48 \pi m_{3/2}^2 M_{\rm P}^2}.
\label{NLSP_decay}
\end{equation}
As for the baryon abundance, it is determined by Eq.(\ref{Yb}) with appropriate value of 
$\varepsilon$ (and $T_{\rm RH}$).

\begin{figure}[h]
\includegraphics[width=90mm]{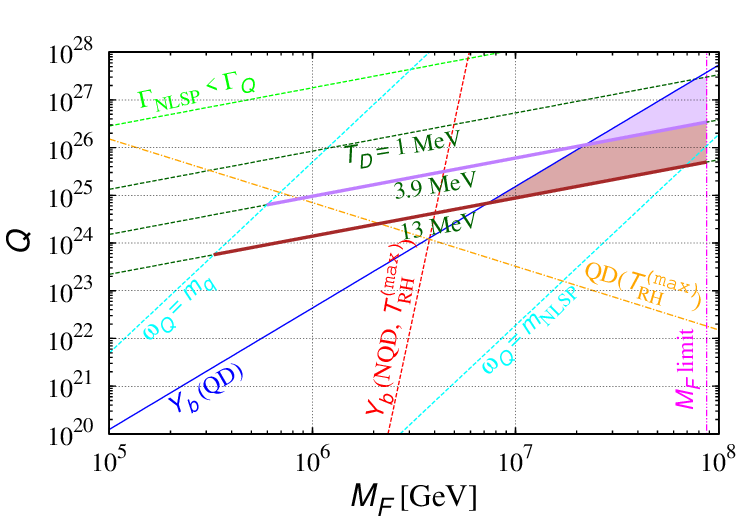}
\caption{Allowed region for $m_{3/2}= 500$ MeV. The NLSP (or equivalently the gravitino 
dark matter) density is fixed by the annihilation (thick solid lines) or determined by the 
$Q$-ball decay (hatched regions). Purple and brown regions correspond to the stau and 
sneutrino NLSPs, respectively.}
\label{fig_indir}
\end{figure}

On the other hand, the latter case appears in relatively narrow range of $T_{\rm D}$.
The argument is similar to the direct production discussed in Sections IV and V. Since the
gravitino abundance has straightforward relation to the baryon abundance through
the number density of the NLSPs as
\begin{equation}
\label{BDM_indir}
5\approx \frac{\rho_{3/2}}{\rho_b} = \frac{m_{3/2}}{m_N}\frac{n_{3/2}}{n_b} \simeq
\frac{m_{3/2}}{m_N} \frac{Q_{\rm cr}}{Q}\frac{B_{\rm NLSP}}{\varepsilon b},
\end{equation}
the baryon abundance is estimated as, from Eq.(\ref{Yb}),
\begin{eqnarray}
\frac{Y_b^{\rm (QD, indir)}}{10^{-10}} & \simeq & 9.5 \times 10^2
\left(\frac{\zeta}{2.5}\right)^{7/2} \left(\frac{N_q}{18}\right)^{-1/2}
\nonumber \\ & & \hspace{-10mm} \times 
\left(\frac{N_{\rm D}}{10.75}\right)^{-1/4}
\left(\frac{m_{\rm NLSP}}{300 \ {\rm GeV}}\right)^{-4}
\left(\frac{m_{3/2}}{20 \ {\rm MeV}}\right)
\nonumber \\ & & \hspace{0mm} \times 
\left(\frac{M_F}{10^7 \ {\rm GeV}}\right)^{7/2}
\left(\frac{Q}{10^{22}}\right)^{-11/8},
\label{Yb_QDindir}
\end{eqnarray}
when the $Q$ balls dominate the energy density of the universe, 
and, for the non $Q$-ball domination case, 
\begin{eqnarray}
\frac{Y_b^{\rm (NQD, indir, s)}}{10^{-10}} & \simeq & 0.42
\left(\frac{\zeta}{2.5}\right)^4 \left(\frac{N_q}{18}\right)^{-1}
\nonumber \\ & & \hspace{-20mm} \times 
\left(\frac{m_{\rm NLSP}}{300 \ {\rm GeV}}\right)^{-4}
\left(\frac{m_{3/2}}{20 \ {\rm MeV}}\right)
\nonumber \\ & & \hspace{-20mm} \times 
\left(\frac{T_{\rm RH}}{10^4 \ {\rm GeV}}\right)
\left(\frac{M_F}{10^7 \ {\rm GeV}}\right)^5
\left(\frac{Q}{10^{22}}\right)^{-1/4},
\label{Yb_NQDindirs}
\end{eqnarray}
for $\omega_Q < m_{\rm NLSP} (= 300$ GeV), and 
\begin{eqnarray}
\frac{Y_b^{\rm (NQD, indir, b)}}{10^{-10}} & \simeq & 0.23
\left(\frac{N_q}{18}\right)^{-1}
\left(\frac{m_{3/2}}{20 \ {\rm MeV}}\right)
\nonumber \\ & & \hspace{-20mm} \times 
\left(\frac{T_{\rm RH}}{10^4 \ {\rm GeV}}\right)
\left(\frac{M_F}{10^7 \ {\rm GeV}}\right)
\left(\frac{Q}{10^{22}}\right)^{3/4},
\label{Yb_NQDindirb}
\end{eqnarray}
for $\omega_Q > m_{\rm NLSP}$. We therefore obtain the relation between $Q$ and $M_F$
from these estimates with $Y_b=10^{-10}$. They are shown as an example of the stau and 
sneutrino NLSP for $m_{3/2}= 500$ MeV in Fig.~\ref{fig_indir}. Thick solid lines denote 
the case when the annihilation fixes the NLSP density, while the $Q$-ball decay determines 
the NLSP density for the hatched regions.

Finally, we comment on the free streaming constraint in the case of the indirect production of 
the gravitino dark matter. Since the decay is determined by the NLSP decay rate (\ref{NLSP_decay}),
the present-day free streaming velocity is estimated as
\begin{equation}
v_0 \simeq \frac{1}{2} \frac{m_{\rm NLSP}}{m_{3/2}}\frac{T_0}{T_{\rm NLSP}}
\left(\frac{N_0}{N_{\rm D}}\right)^{1/3},
\end{equation}
where $T_{\rm NLSP} = (90/4\pi^2N_{\rm D})^{1/4}\sqrt{\Gamma_{\rm NLPS} M_{\rm P}}$. This turns 
out the constraint on the NLPS mass as $m_{\rm NLSP} \gtrsim 300$ GeV.



\end{document}